# Polaronic correlations and phonon renormalization in $La_{1-x}Sr_xMnO_3$ ($x = 0.2, 0.3$)

M. Maschek[1,2], J.-P. Castellan[1,3], D. Lamago[1,3], D. Reznik[4] and F. Weber[1]

[1] *Institute for Solid State Physics, Karlsruhe Institute of Technology, D-76021 Karlsruhe, Germany*

[2] *Fundamental Aspects of Materials and Energy, Faculty of Applied Sciences, Delft University of Technology, Mekelweg 15, 2629 JB Delft, The Netherlands*

[3] *Laboratoire Léon Brillouin (CEA-CNRS), CEA-Saclay, F-91911 Gif-sur-Yvette, France*

[4] *Department of Physics, University of Colorado at Boulder, Boulder, Colorado, 80309, USA*

According to standard theory the magnetoresistance magnitude in ferromagnetic manganites crucially depends on the electron-phonon coupling strength. We showed that in $La_{0.7}Sr_{0.3}MnO_3$ the phonon renormalization is strong, despite its relatively small magnetoresistance. Here, we report results of a similar inelastic neutron scattering investigation of a closely related compound, $La_{0.8}Sr_{0.2}MnO_3$, where the magnetoresistance is enhanced. We find similar phonon renormalization and dynamic CE-type polaron correlations as in $La_{0.7}Sr_{0.3}MnO_3$. However, quantitative comparison of the results for the two samples shows that only polaron lifetime is well correlated with the strength of the CMR.

## I. INTRODUCTION

The manganites have complex phase diagrams, due to strongly competing magnetic, orbital, charge and lattice degrees of freedom [1]. This competition leads to a unique property in the ferromagnetic manganites known as colossal magnetoresistance (CMR) [2,3]. The CMR manganites exhibit a simultaneous transition from a ferromagnetic (FM) and metallic ground state to a paramagnetic (PM) insulating phase at elevated temperatures. An applied magnetic field stabilizes ferromagnetism and results in a strongly reduced resistivity at the zero-field Curie temperature $T_C$. In the insulating phase strong electron-lattice coupling via the Jahn-Teller (JT) [4] effect favors lattice distortions that trap the charge carriers leading to the formation of polarons [5-7]. These polarons involve cooperative lattice distortions that correspond to short-range charge and orbital order (COO) of $Mn^{3+}$ (JT active) and $Mn^{4+}$ (JT inactive) ions [8,9]. In many manganites correlations of those polarons result in superstructures of the CE-type [10,11] on heating through $T_C$ into the PM phase.

The CE-type short range COO associated with the ordering wave vector $\mathbf{q}_{CE} \approx (1/4, 1/4, 0)$ [10] forms in prototypical CMR manganites such as $La_{0.7}Ca_{0.3}MnO_3$ [12] and $La_{1.2}Sr_{1.8}Mn_2O_7$ [13].

Magnitude of the magnetoresistance effect is expressed by $-(\rho(B) - \rho(0))/\rho(0)$ [$\rho(B)$: resisitivity at magnetic field $B$]. Many CMR manganites have an inverse correlation between magnetoresistance magnitude and $T_C$: The smaller the $T_C$ the bigger the resistivity jump on applying magnetic field [1]. Standard theory [5,6] links the absolute value of $T_C$ directly to the relative strengths of the competing interactions: A low $T_C$ indicates strong electron-phonon coupling (EPC), promoting JT distortions, whereas a high $T_C$ points to weak EPC.

Here we focus on the FM manganites of the $La_{1-x}Sr_xMnO_3$ family have a rhombohedral structure below 750 K. The Curie temperatures in the Sr-doped system are among the highest of all FM manganites with $T_C = 350$ K and 305 K for x = 0.3 and 0.2, respectively [14,15].

Previously we investigated $La_{0.7}Sr_{0.3}MnO_3$ [16], which had been considered to be a pure double-exchange system with minimal JT electron-lattice interaction. However, we found that this compound also features CE-type correlated polarons and strong electron-phonon renormalization. The polarons correlations are short-range and dynamic [16] with a correlation length for the polaronic fluctuations of $\xi_{x=0.3} = 34(4)$ Å and the life time of 1.00(15) ps. Similar results have been reported by others [17,18].

The magnetoresistance at $T_C$ in $La_{0.7}Sr_{0.3}MnO_3$ is 0.35 for $B = 15$ T. Here we focus on the slightly lower doped $La_{0.8}Sr_{0.2}MnO_3$ where the FM-PM transition is accompanied by a metal-insulator transition. This subtle difference in the phase transition is reflected in much stronger CMR, which is 0.75 for 15T in $La_{0.8}Sr_{0.2}MnO_3$ [14]. This contrast should be reflected in its atomic lattice response (both in the phonons and the polaronic distortions) compared to that of $La_{0.7}Sr_{0.3}MnO_3$.

We performed inelastic neutron scattering measurements of $La_{0.8}Sr_{0.2}MnO_3$ focusing on the evolution of the transverse acoustic (TA) phonon mode propagating along the [110] direction and scattering from correlated polarons in the vicinity of $T_C$. Then we compared the results with our previous investigation of $La_{0.7}Sr_{0.3}MnO_3$ reported in Ref. [16].

## II. EXPERIMENT

The sample was a high-quality single crystal of $La_{0.8}Sr_{0.2}MnO_3$ with a volume of about 0.7 cm$^3$. Inelastic neutron measurements were performed on the 1T neutron triple-axis spectrometer at the ORPHEE reactor (LLB, CEA Saclay) using doubly focusing PG002 monochromator and analyzer crystals. The final energy was fixed to 14.7 meV. We performed energy scans at a constant momentum-transfer $Q = \tau + q$ where $\tau$ and $q$ are reciprocal lattice vector and reduced momentum transfer, respectively. The experimental resolution was obtained from standard calculations using the "*rescal*" program package [19]. In our work on $La_{0.8}Sr_{0.2}MnO_3$ we used the same experimental conditions as in our work on $La_{0.7}Sr_{0.3}MnO_3$ [16] in order to be able to compare results at different Sr doping.

The perovskite manganites have an ideal cubic structure ($Pm\bar{3}m$) above ~ 750 K [1]. With cooling $La_{1-x}Sr_xMnO_3$ acquires a rhombohedral structure ($R\bar{3}c$), as a result of a rotation of the $MnO_6$ octahedra around the [111] axis. $La_{0.8}Sr_{0.2}MnO_3$ shows another structural transition to a orthorhombic phase below $T \approx 120$ K [1]. However, it was shown that low-energy phonons, in particular acoustic modes, can still be well described by a cubic shell model [20]. Therefore, we use the cubic notation for all wave vectors. The wave vectors are given in reciprocal lattice units (r.l.u.) of ($2\pi/a$, $2\pi/b$, $2\pi/c$), where $a = b = c = 3.86$ Å.

## III. RESULTS

We investigated [1$\bar{1}$0]-polarized TA phonons of $\Sigma_3$-symmetry dispersing in the [110] direction at various temperatures. The symmetry of $\Sigma_3$-phonons match the displacement pattern of the JT distortion of the $MnO_6$-octahedra and the COO of the CE-type [20,21]. The latter is characterized by superlattice peaks at $q_{CE} = (1/4, 1/4, 0)$, e.g., in $La_{0.7}Ca_{0.3}MnO_3$ [9] and $La_{1.2}Sr_{1.8}Mn_2O_7$ [22]. The structure factor of these TA phonons is large close to the reciprocal lattice vector $\tau = (2, 2, 0)$, where measurements can be performed in a purely transverse geometry, i.e., $\tau \perp q$ in $Q = \tau + q = (2, 2, 0) + (-h, +h, 0)$.

The neutron scattering intensities at $Q = (2-h, 2+h, 0)$ as a function of $h$ are represented in color-coded contour plots shown in Figs. 1(a)-(c). We observe sharp phonon peaks and a well-defined dispersion at low temperatures [Fig. 1(a)]. The TA phonon branch dominates the spectrum at $h \leq 0.3$. There is an anti-crossing of the TA branch with a transverse optical (TO) branch near $h \approx 0.35$. Close to the zone boundary ($h \approx 0.45 - 0.5$) two well-separated phonons appear at 8.3 meV and 15.8 meV.

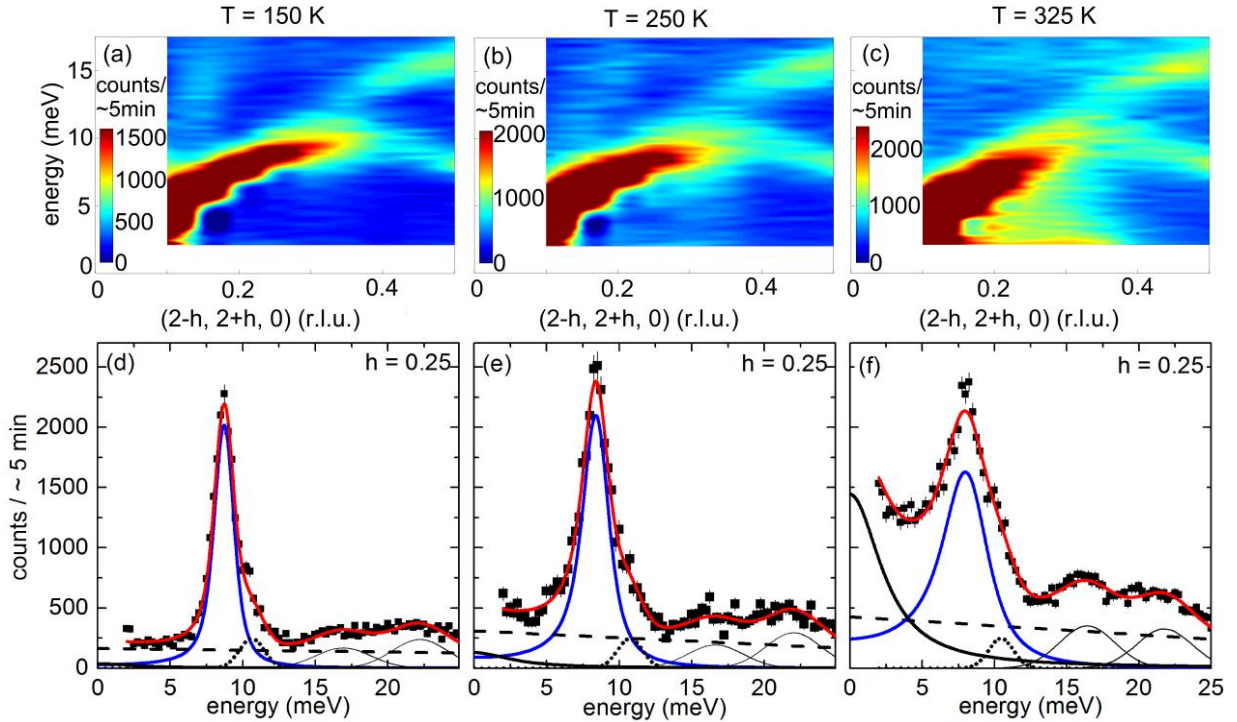

**Figure 1**: (a)-(c) Color-coded contour maps representing inelastic neutron scattering intensities along $Q = (2-h, 2+h, 0)$, $0.1 \leq h \leq 0.5$ and energies 2 meV $\leq E \leq$ 17 meV. (d)-(f) Corresponding energy scans at $Q = (1.75, 2.25, 0)$ for $T = 150$ K (d), 250 K (e) and 325 K (f). In (d)-(f), solid (red) lines are fits consisting of a damped harmonic oscillator function for the TA mode (solid & blue), a Lorentzian for quasielastic scattering (solid & black, see text) and the estimated background (dashed). The Gaussians (thin black lines) denote other modes and the dotted lines denote temperature independent artefacts (E ≈ 10 meV).

We studied the wave-vector and temperature dependence of the TA and TO modes by analyzing energy scans at constant momentum transfer as shown in Figs. 1(d)-(f). For $E \leq 5$ meV the experimental background for each temperature was obtained from the zone boundary scans. For $E \geq 12$ meV it was obtained from the zone center (not shown) and approximated by a straight line [dashed lines in Figs. 1(d)-1(f)] for the energies in between.

Energy scans at constant-momentum transfers were approximated with damped harmonic oscillator (DHO) functions convoluted with the calculated Gaussian resolution [Figs. 1(d)-1(f)]. The Bose factor built in to the DHO function accounts for the observed integrated intensities of the TA phonons at the zone boundary at increasing temperatures. The peak just above 10 meV is an artifact exhibiting no detectable temperature dependence. Hence, it was described by a temperature-independent Gaussian function and posed no problem for further analysis [16]. Higher-energy TO phonons were fitted with simple Gaussians. A reduced phonon energy can be generally expected on heating because of thermal expansion and increasing thermal atomic motions. However, such a behavior should be similar for all phonon modes. The TA zone boundary phonon at $h = 0.5$ shows an energy reduction (softening) of 3 % on heating from 150 K to 325 K and its intensity increases according to the Bose factor. On the other hand, the TA mode at $h = 0.25$ softens by 0.81 meV, i.e., 9% and acquires an intrinsic line width of 1.2 meV on heating from 150 K to 325 K [Figs. 1(d)-(f)]. Furthermore, the intensity increase between 250 K [Fig. 1(e)] and 325 K at low energies [Fig. 1(f)] cannot be explained alone by the TA phonon mode. Hence, we introduced a Lorentzian peak centered at $E = 0$ describing quasielastic (QE) scattering if the line width is not resolution limited.

Heating through $T_C = 305$ K leads to a pronounced low-energy tail of the TA phonon around $\boldsymbol{q}_{CE} = (1/4, 1/4, 0)$ [Fig. 1(c)]. Detailed temperature dependences of phonon softening ($\Delta E$) and line width for $h = 0.15, 0.25$ and $0.5$ are shown in Figs. 2(a)(b). We find similarly strong softening at the two smaller wave vectors whereas the broadening is strongest close to $h = 0.25$, which corresponds to the COO ordering wave vector $\boldsymbol{q}_{CE} = (1/4, 1/4, 0)$ observed in many FM manganites. Note that softening and broadening expected from thermal expansion and thermally induced disorder are considerably smaller as observed in the results for the zone boundary mode at $h = 0.5$ [Figs. 2(a),(b)].

The anomalous softening and broadening extend up to about the Curie temperature of 305 K, which we determined from the temperature dependence of the (110) Bragg reflection [Fig. 2(c)]. At higher temperatures, further softening becomes consistent with thermal expansion effects also observed at $h = 0.5$. The intensity of the Lorentzian at $E = 0$ and $h = 0.25$ strongly increases within the same temperature range as the phonon renormalization and reaches a maximum close to $T_C$ [Fig. 2(d)].

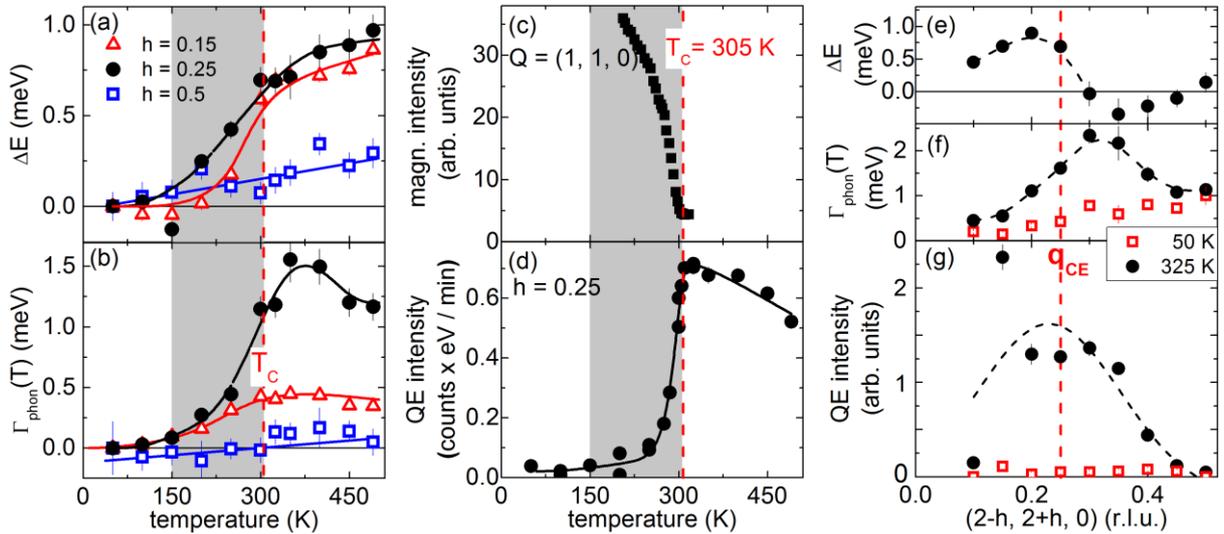

**Figure 2**: (left) Temperature dependences of (a) phonon softening $\Delta E$ and (b) phonon line width $\Gamma_{phon}$ (HWHM) with respect to results obtained at $T = 50$ K data shown for $\boldsymbol{Q} = (2-h, 2+h, 0)$, $h = 0.15, 0.25$ and $0.5$. (middle) Temperature dependences of (c) the intensity of the (110) Bragg peak with a strong ferromagnetic contribution below the Curie temperature $T_C = 305$ K and (d) quasielastic scattering intensity at $h = 0.25$ extracted from raw data as shown in Figs. 1 and 3. Solid lines are guides to the eye. The temperature range characterized by anomalous phonon softening and broadening at $h = 0.15$ and $0.25$ is marked in grey in panels (a)-(d). The upper end of the grey region coincides with the Curie temperature $T_C = 305$ K. (right) Momentum dependences of phonon (e) softening $\Delta E$, (f) line width $\Gamma_{phon}$ and (g) QE intensity at $T = 325$ K (dots) compared to results obtained at $T = 50$ K (open squares). Dashed lines are Gaussian fits to the data at $T = 325$ K over the indicated momentum ranges.

The momentum dependences observed at 325 K ($> T_C$) show clear maxima along the [110] direction [Figs. 2(e)(f)(g)]. The maxima for the softening, broadening, and QE intensity are at $h = 0.19$, 0.3 and 0.23, respectively [see dashed lines in Figs. 2(e)(f)(g)]. This behavior is similar but less pronounced in the sample with $x = 0.3$ [16]. The width in momentum space of QE scattering in $La_{1-x}Sr_xMnO_3$ with $x = 0.2$ seems to be slightly larger than for $x = 0.3$ giving a correlation length of $\xi_{x=0.2} = 28(6)$ Å vs. $\xi_{x=0.3} = 34(4)$ Å. However, the difference depends entirely on the outlier at $h = 0.15$, which is possibly related to an artefact produced by the resolution function combined with the proximity to a strong Bragg peak for small $h$. Unfortunately, the scattering angle in the experiment limited further investigations at smaller energies.

Thus our data reveal that phonon renormalization along with QE scattering intensity are strongest at $T_C$. [Figs. 2(e)-(g)]. QE scattering intensity is strongest close to $q_{CE}$ and the largest phonon softening and broadening are located just below and above this wave vector respectively. Hence, we focus on the temperature dependences deduced from measurements at $q_{CE}$ in the following.

As in our previous investigation of $La_{0.7}Sr_{0.3}MnO_3$ [16], we assign the strong QE Lorentzian scattering to dynamic polaron correlations of the CE-type, which effectively trap the conduction electrons above $T_C$. However, the energy scans shown in Figs. 1(d)-(f) were not adequate to determine the line width of the Lorentzian peak, i.e., the life time of the QE scattering. In order to determine both amplitude and line width of the Lorentzian at $h = 0.25$, we performed additional INS measurements for $La_{1-x}Sr_xMnO_3$ with $x = 0.2$ in a slightly different configuration of the instrument, which allowed energy transfers from -4 to +15 meV. In these scans, the QE peak, which strongly increases at $T_C$, appears beneath a resolution-limited and practically temperature-independent elastic line [Fig. 3(a)-(d)].

After correcting for the experimental resolution, the average width of the approximated Lorentzians at $300\ K \leq T \leq 325$ is $\Gamma_{polaron} = (2.03 \pm 0.23)$ meV [half width at half maximum (HWHM)], which corresponds to the lifetime of the polaron correlations of $t = (2.04 \pm 0.23)$ ps. By the same procedure we obtain $\Gamma_{polaron} = (4.15 \pm 0.63)$ meV (HWHM) and $t = (1.0 \pm 0.15)$ ps at $325\ K \leq T \leq 400$ K for $La_{0.7}Sr_{0.3}MnO_3$ [16]. The temperature dependence of QE scattering line width and the deducted lifetime in $La_{1-x}Sr_xMnO_3$ for $x = 0.2$, 0.3 are given in Figs. 4(a)(b). Although we detected the QE peak well below 300 K, its width could only be determined with high accuracy closely below and above $T_C$ where the QE intensity is greatly enhanced. Uncertainty in the line width increases strongly on cooling well below $T_C$ in both compounds, due to strongly decreasing QE scattering intensities.

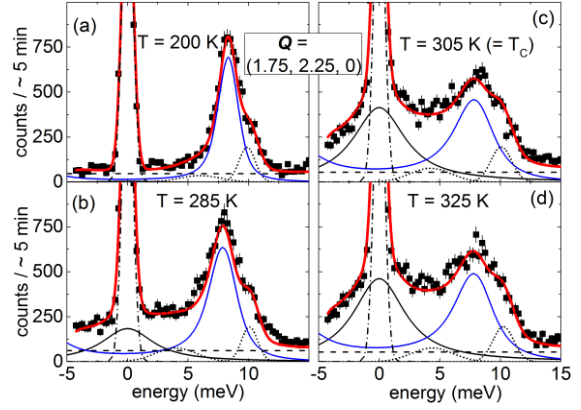

**Figure 3**: Energy scans from -4 meV to 15 meV at $Q = (1.75, 2.25, 0)$ for (a) $T = 200$ K, (b) 285 K, (c) 305 K and (d) 325 K. Solid (red) lines are fits consisting of a damped harmonic oscillator function for the TA mode (solid & blue), a Lorentzian for QE scattering (solid & black) and the estimated background (dashed). The dotted lines denote temperature independent artefacts (~ 5 and 10 meV). The resolution-limited line at $E = 0$ (dashed-dotted line) is due to incoherent scattering and is practically temperature independent.

QE scattering intensities at the two doping levels are harder to compare than peak positions and line widths, because they depended on sample size. We employed the intensities of the TA phonon at $h = 0.25$ and $T = 200$ K measured in both compounds [Fig. 5] in order to calculate the scaling factor between the neutron scattering data taken on the two samples. We determine it at 200 K because the influence of the QE scattering is still negligible and the background underneath the phonon is flat. The analysis in Figure 5 shows that there is a factor of 0.33 difference in the absolute intensities in the two measurements, i.e., the scattering intensities obtained for the $La_{1-x}Sr_xMnO_3$ sample with $x = 0.3$ can be scaled to match those obtained for $x = 0.2$ if they are multiplied by 0.33 [Fig. 5].

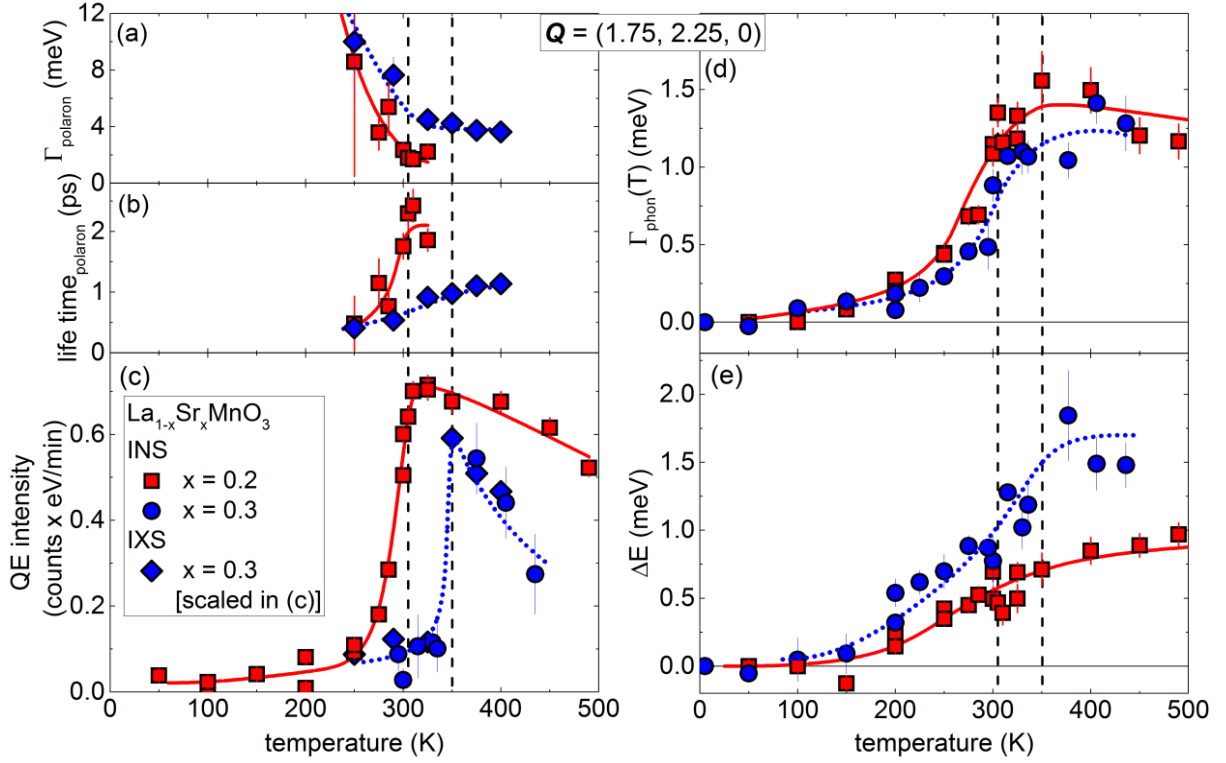

**Figure 4**: QE scattering (left) and phonon renormalization (right) for $La_{1-x}Sr_xMnO_3$ (x = 0.2, 0.3) at h = 0.25. (a) shows line width $\Gamma_{polaron}$, (b) lifetime and (c) normalized intensity of QE scattering associated with CE polaronic correlations. (d) shows phonon line width $\Gamma_{phon}$ (HWHM) and (e) Difference between the phonon energy at low temperature and just above $T_C$, $\Delta E$. The intensities of the x = 0.3 data in (c) were normalized to x = 0.2 data by multiplication with a factor of 0.33 (see text and Fig. 5). Solid and dotted lines are guides to the eye. Vertical dashed lines represent $T_C$ = 305 K and 350 K for x = 0.2, 0.3.

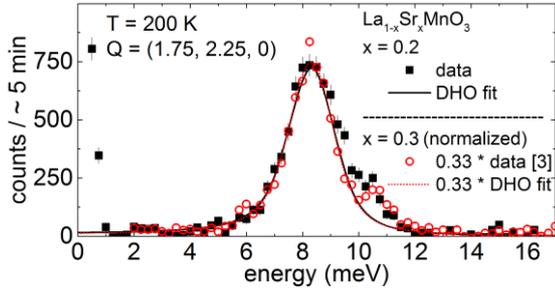

**Figure 5**: (a) Energy scan (background subtracted) of the TA phonon at $Q$ = (1.75, 2.25, 0) and $T$ = 200 K for $La_{1-x}Sr_xMnO_3$ (x = 0.2, 0.3). The intensities of x = 0.3 data can be normalized to those of x = 0.2 data by multiplication with a factor of 0.33 obtained from the ratio of integrated intensities. The small peak at 10.5 meV is a temperature independent artifact.

Our comparison for the two different doping levels shows that QE scattering intensity is higher in the lower-doped sample featuring a larger CMR effect [Fig. 4(c)]. Yet the average increase from x = 0.3 to x = 0.2 is about a factor of 1.5, i.e., smaller than the factor of two reported for the CMR strength (see table 1) [14]. In contrast, phonon broadening is the same [Fig. 4(d)] and the phonon softening is smaller [Fig. 4(e)] for the TA mode at h = 0.25 in $La_{0.8}Sr_{0.2}MnO_3$. Only the difference of the life times of the polaronic correlations, 2.04(23) ps and 1.00(15) ps for $La_{0.8}Sr_{0.2}MnO_3$ and $La_{0.7}Sr_{0.3}MnO_3$, respectively [Fig. 4(b)], match the difference in the CMR strength.

## IV. DISCUSSION

Table 1 summarizes our results together with available analogous results on other CMR manganites taken from literature. We clearly see that the increase of the CMR strength is correlated with a decreasing Curie temperature. However, it is the main result of our investigation that this increase of the CMR strength is, at least in the $La_{1-x}Sr_xMnO_3$ family, not related to a corresponding increase of EPC since the increase of the line widths of the TA phonon across $T_C$ are the same in $La_{0.7}Sr_{0.3}MnO_3$ and $La_{0.8}Sr_{0.2}MnO_3$. Instead, our results indicate that the strength of the CMR scales (within uncertainty) with polaron lifetime.

The polaronic lattice distortion magnitude, which determines the intensity of the quasielastic scattering in our data can be expected to have a strong effect on the strength of the CMR. However, we show that it seems to be similar in $La_{0.8}Sr_{0.2}MnO_3$ compared to $La_{0.7}Sr_{0.3}MnO_3$. Furthermore, all available data indicate that these intensities are of the same order of magnitude [23] even including the bilayer manganite $La_{1.2}Sr_{1.8}Mn_2O_7$ [13,21] and $La_{0.7}Ca_{0.3}MnO_3$ [12], where CMR effects are up to two orders of magnitude stronger than in $La_{1-x}Sr_xMnO_3$.

| material | $T_C$ (K) | $B$ (T) | $\frac{-(\rho(B)-\rho(0))}{\rho(0)}$ | $\frac{\rho(0)}{\rho(B)}@T_C$ | Pol. life time (ps) | Corr. length $\xi$ (Å) | Reference |
|---|---|---|---|---|---|---|---|
| $La_{0.7}Sr_{0.3}MnO_3$ | 350 | 15 | 0.35 | 2 | 1.00(15) | 34(4) | [18] |
| $La_{0.8}Sr_{0.2}MnO_3$ | 305 | 15 | 0.75 | 4 | 2(23) | 28(6) | [18] |
| $La_{0.8}Ca_{0.2}MnO_3$ | 178 | - | - | - | > 2 | 14 | [1] |
| $La_{0.75}Ca_{0.25}MnO_3$ | 245 | 4 | 0.84 | 6 | - | - | [13] |
| $La_{0.7}Ca_{0.3}MnO_3$ | 227 | 5 | 0.97 | 30 | > 2 | 28 | [1],[24] |
| $La_{1.2}Sr_{1.8}Mn_2O_7$ | 118 | 14 | 0.99 | 300 | - | 16 | [22] |

**Table 1**: Curie temperatures, applied magnetic field and corresponding magnetoresistance $\frac{-(\rho(B)-\rho(0))}{\rho(0)}$ and drop of resistivity at $T_C$ ($\frac{\rho(B=0)}{\rho(B\neq 0)}$), polaron life time and correlation length of some CMR compounds. The polaron life time measured in $La_{1-x}Ca_xMnO_3$ ($x$ = 0.2, 0.3) is limited to 2 ps, set by the energy resolution of spectrometer [1].

Similarly, correlation lengths $\xi$ of polarons are of the same order of magnitude for several CMR compounds listed in table 1, despite strongly varying strengths of the CMR effects. Therefore, it is unlikely that the correlation length is the driving factor for the CMR.

Incipient structural phase transitions involving short-range polaronic fluctuations tend to destabilize the atomic lattice leading to softening of phonons resembling the CE-type charge and orbital order. If the life time of polaronic fluctuations increases, the lattice should stabilize into a new structure, which is expected to result in smaller phonon softening. We observe this behavior in $La_{1-x}Sr_xMnO_3$, where the $x$ = 0.2 sample shows a larger polaron life time resulting in harder phonons, i.e a smaller phonon softening compared to $x$ = 0.3.

Strong phonon renormalization effects indicating QE fluctuations in the FM phase of $La_{1.2}Sr_{1.8}Mn_2O_7$ develop above $T_C$ into the characteristic Jahn-Teller distortion with an elastic superlattice peak close to $q_{CE}$ [21], where the CE-order is short-range and static. Another investigation of the bilayer manganite $La_{1.2}Sr_{1.8}Mn_2O_7$ by [22] shows, that the line width of QE and elastic scattering slightly above $T_C$ is resolution limited ($\approx$2 meV) and not measurable below $T_C$. However, the presence of an elastic superlattice peak in $La_{1.2}Sr_{1.8}Mn_2O_7$ indicates significantly larger lifetimes of the polarons compared to our results in $La_{1-x}Sr_xMnO_3$. Thus a quantitative analysis of polaron life time below $T_C$ is very desirable for further investigations in $La_{1.2}Sr_{1.8}Mn_2O_7$, due to its significantly larger CMR effect compared to $La_{1-x}Sr_xMnO_3$.

Another example resembling the behavior described before is observed in the 50% doped bilayer manganite $LaSr_2Mn_2O_7$, which however has an insulating low-temperature ground state exhibiting a long-range charge and orbital CE order [24] with no metal-insulator transition. This polaronic ground state is comparable to the high temperature insulating phase above $T_C$ in $La_{1-x}Sr_xMnO_3$ ($x$ = 0.2, 0.3) or $La_{1.2}Sr_{1.8}Mn_2O_7$. The difference compared to e.g. $La_{1-x}Sr_xMnO_3$ ($x$ = 0.2, 0.3) is, that the half-doped $LaSr_2Mn_2O_7$ shows a "perfect" ordering of $Mn^{3+}$ (JT active) and $Mn^{4+}$ (JT inactive) ions leading to a static polaronic lattice with an "infinite" life time. Within the COO state the transverse acoustic phonon at $q_{CE}$ [see Fig 3(b) in [24]], shows no renormalization effects in agreement with our proposition, i.e., that an infinite life time of the polaronic correlations results in a very small or absent phonon renormalization.

In comparison to $La_{1-x}Sr_xMnO_3$ and in respect to CMR strength (table 1), we expect polaron life times that are one order of magnitude ($\approx$ 10 ps) larger in $La_{1-x}Ca_xMnO_3$ (0.2 – 0.3) and two orders of magnitude ($\approx$ 100 ps) larger in the bilayer manganite $La_{1.2}Sr_{1.8}Mn_2O_7$. A detailed comparative high energy-resolution study of phonons, intensities and life times of polarononic correlations in, e.g., the bilayer manganite $La_{2-x}Sr_{1+2x}Mn_2O_7$ and the $La_{1-x}Ca_xMnO_3$ series would be highly desirable in order to investigate these relations over a larger range in the strength of the CMR effect.

## V. CONCLUSION

We report inelastic neutron scattering data revealing dynamic polaronic correlations of CE-type in $La_{0.8}Sr_{0.2}MnO_3$ similar to our previous findings in higher-doped $La_{0.7}Sr_{0.3}MnO_3$ [16]. Strong renormalization of the TA phonons propagating along the [110] is associated with the polaronic behavior. A quantitative comparison of phonons and polarons in $La_{1-x}Sr_xMnO_3$, $x$ = 0.2 and 0.3, shows that lifetime and intensity of polaronic correlations scale with the strength of CMR but phonon renormalization and other properties, e.g., polaronic correlation length do not. This study needs to be extended to compounds with larger CMR (see Table 1) in order to test the proposed correlation between the CMR strength and the life time of polaronic correlations in FM manganites.

## ACKNOWLEDGMENTS

M.M. and F. W. were supported by the Helmholtz Society under the Contract No. VH-NG-840. D.R. was supported by the DOE, Office of Basic Energy Sciences, Office of Science, under Contract No. DE-SC0006939